\documentclass[sigconf]{sig-alternate}
\usepackage{amsmath,epsfig}
\usepackage{booktabs} 
\usepackage{graphicx}
\usepackage{svg}
\usepackage{epstopdf}
\usepackage{verbatim}
\usepackage{lipsum}
\usepackage{mathtools}
\usepackage{float}
\usepackage{multirow} 
\usepackage{import}
\usepackage{tabu}
\usepackage[export]{adjustbox}
\usepackage{quoting}
\usepackage{array}
\usepackage{tabularx}
\usepackage{listings}
\usepackage{color}
\usepackage{frcursive}
\usepackage{calligra}
\usepackage[T1]{fontenc}

\usepackage{balance}
\usepackage{listings}

\def \ElasticScaling       {{\bf [ElasticScaling]}}
\def \DatacenterAMP        {{\bf [DatacenterAMP]}}
\def \GPU                  {{\bf [GPU]}}
\def \SoftErrors           {{\bf [SoftErrors]}}
\def \InformationDiffusion {{\bf [InformationDiffusion]}}
\def \DatabaseScalability  {{\bf [DatabaseScalability]}}
\def \TransactionalMemory  {{\bf [TransactionalMemory]}}
\def \CloudTransactions    {{\bf [CloudTransactions]}}
\def \RouterBuffer         {{\bf [RouterBuffer]}}
\def \MapReduce            {{\bf [MapReduce]}}
\def \PPVoD                {{\bf [P2PVoD]}}
\def \Roofline             {{\bf [Roofline]}}
\def \WiFi                 {{\bf [802.11]}}
\def \TCP                  {{\bf [TCP]}}
\def \fiveG                {{\bf [5G]}}
\def \Tay                  {{\bf [T]}}
\def \TGS                  {{\bf [TGS]}}
\def \HM                   {{\bf [HM]}}

\begin{document}
\title{A review of analytical performance modeling \\
       and its role in computer engineering and science
\thanks{Some parts of this article appeared in [T] and
``Lessons from Teaching Analytical Performance Modeling'',
{\it Proc. ICPE Workshop on Education and Practice of Performance Engineering},
Mumbai, India (April 2019), 79--84.
Many thanks to Haifeng Yu and Prashant Shenoy for reading the draft 
and making many helpful suggestions.
}
}
\author{
 \alignauthor{Y.C. Tay}\\
       \affaddr{National University of Singapore}\\
       \email{dcstayyc@nus.edu.sg}
}

\maketitle

\begin{abstract}

This article is a review of 
{\sl analytical performance modeling for computer systems}.
It discusses the motivation for this area of research, 
examines key issues, introduces some ideas, illustrates how it is applied, 
and points out a role that it can play in developing Computer Science.

\end{abstract}

\section{Introduction}
\label{sec:Intro}

A {\it computer system} has multiple components, 
possibly a mix of both hardware and software.
It may be a processor architecture on a chip, 
or a software suite that supports some application in the cloud.
In our context, analytical modeling essentially refers to 
the formulation of equations to describe the performance of the system.

System performance can be measured by some metric $y$,
say throughput, latency, availabilty, etc.
This $y$ depends on the workload on the system,
such as the number of concurrent threads running on a chip,
or the number of users for an application.
It also depends on the system configuration,
like the number of cores in a processor,
or the number of virtual machines processing user requests.
Collectively, we refer to the numerical workload and configuration variables 
as input parameters $\bf x$.

Abstractly speaking, a computer system determines a function $f$
that maps input parameters $\bf x$ to the perfomance metric $y$, 
i.e. $y=f({\bf x})$.
The complexity and nondeterminism in a real system makes 
it impossible to know this $f$ exactly.
The purpose of an analytical model is then to derive a function $\hat{f}$
that approximates the ground truth $f$,
i.e. $\hat{f}({\bf x})\approx y$.

There are various ways of constructing $\hat{f}$.
Using a statistical approach, one might start with a sample of 
$\langle {\bf x}, y\rangle$ values, fix a class of functions for $\hat{f}$,
and determine the $\hat{f}$ that best fits the sample according to some
optimization criterion (mean square error, say).
Alternatively, a machine learning approach might
use the sample to train an artificial neural network that computes $\hat{f}$.
In both cases, $\hat{f}$ is determined from 
the sample $\langle {\bf x}, y\rangle$,
so no knowledge of the system is used.
It follows that we may not be able to use $\hat{f}$ to analyze
the behavior of the system, 
to understand why one particular hardware architecture is better than
another, or what combination of parameters can cause performance to 
deteriorate, etc.
Such an $\hat{f}$ is thus a {\bf blackbox} model of $f$.

Sometimes, very little is known about the system itself:
the processor architecture may be proprietary,
or the server cluster may be virtualized.
It such cases, it is entirely appropriate 
that we resort to a blackbox model of the system.

In many cases, the performance analyst may actually have a sufficiently detailed
understanding of the system to mathematically describe the relationships
among the speeds and delays within it.
One can then use equations to {\it construct} $\hat{f}$.
Such an $\hat{f}$ is then a {\bf whitebox} model:
it is based on a mathematical analysis of the system,
can be analyzed by standard mathematical techniques
(calculus, probability, etc.),
and the analysis can be interpreted in terms of details in the system.
Therein lies the power of an analytical model.

Our review begins in Sec.~\ref{sec:Why}
by describing a common motivation in constructing
an analytical model, namely to engineer a computer system:
to predict the throughput for a workload, 
to determine how failures affect response time,
to evaluate the performance implication for different design choices, etc.

The derivations in an analytical model are often based on strong assumptions;
this is often considered a weakness of the approach.
Sec.~\ref{sec:Assumptions} presents examples to show 
that these models are often accurate 
even when the assumptions are violated. 
In addition to the assumptions, the derivations may make approximations
that are hard to justify theoretically (Sec.~\ref{sec:AVA}).
However, the final arbiter for whether the assumptions and approximations
are acceptable lies not in theory, but in the experimental validation
of the model.

One broadly applicable technique in modeling is bottleneck analysis 
(Sec.~\ref{sec:Bottleneck}).
It is powerful in that it requires very little information about
the details in a system, but it is also weak in that it offers only
performance bounds.
Nonetheless, these may suffice for some purposes, 
like comparing scalability limits, say.

A crucial advantage that analytical models have over simulation models
lies in the global view of the parameter space that they offer.
One can often use the equations to identify important (and unimportant)
regions of the space, reduce the number of parameters, etc. 
(Sec.~\ref{sec:Parameters}). 

A computer system can have many components that interact in complicated ways,
but Sec.~\ref{sec:Decomposition} shows how this complex interaction 
can be decomposed into separate smaller models.  
Such a decomposition then provides a way
to study the separate impact of hardware and software on a workload,
the interaction between resource and data contention, etc.

Since an analytical model is an approximation of the real system,
we must verify not just the accuracy of its numerical predictions,
but also check that crucial properties revealed by the model 
(are not just artefacts of the model, but)
are in fact properties of the real system.
Sec.~\ref{sec:AnalyticValidation} discusses this concept of analytic validation.

In contrast to the engineering motivation for an analytical model,
Sec.~\ref{sec:Analysis} points out the role that such models can play 
in developing a science for the behavior of these engineered systems.

Throughout, we will draw examples from recent literature on hardware,
networking and datacenters, etc. to show the use of analytical models
in designing, controlling and studying systems, large and small. 
The range of examples is wide, 
as wide as the broad perspective that we want our students to have.

Inevitably, we will need to use notation (e.g. $M/D/1$) and 
terminology (e.g. availability), 
and mention techniques (e.g. MVA) and results (e.g. Little's Law)
from performance analysis.
However, the reader can look these up easily with a web search,
and no in-depth technical or theoretical knowledge 
is required for understanding the review.

Sec.~\ref{sec:Conclusion} concludes with a summary of the review.

\section{Why an analytical model?}
\label{sec:Why}

\begin{figure}[t]
\centering
\includegraphics[height=5.9cm]{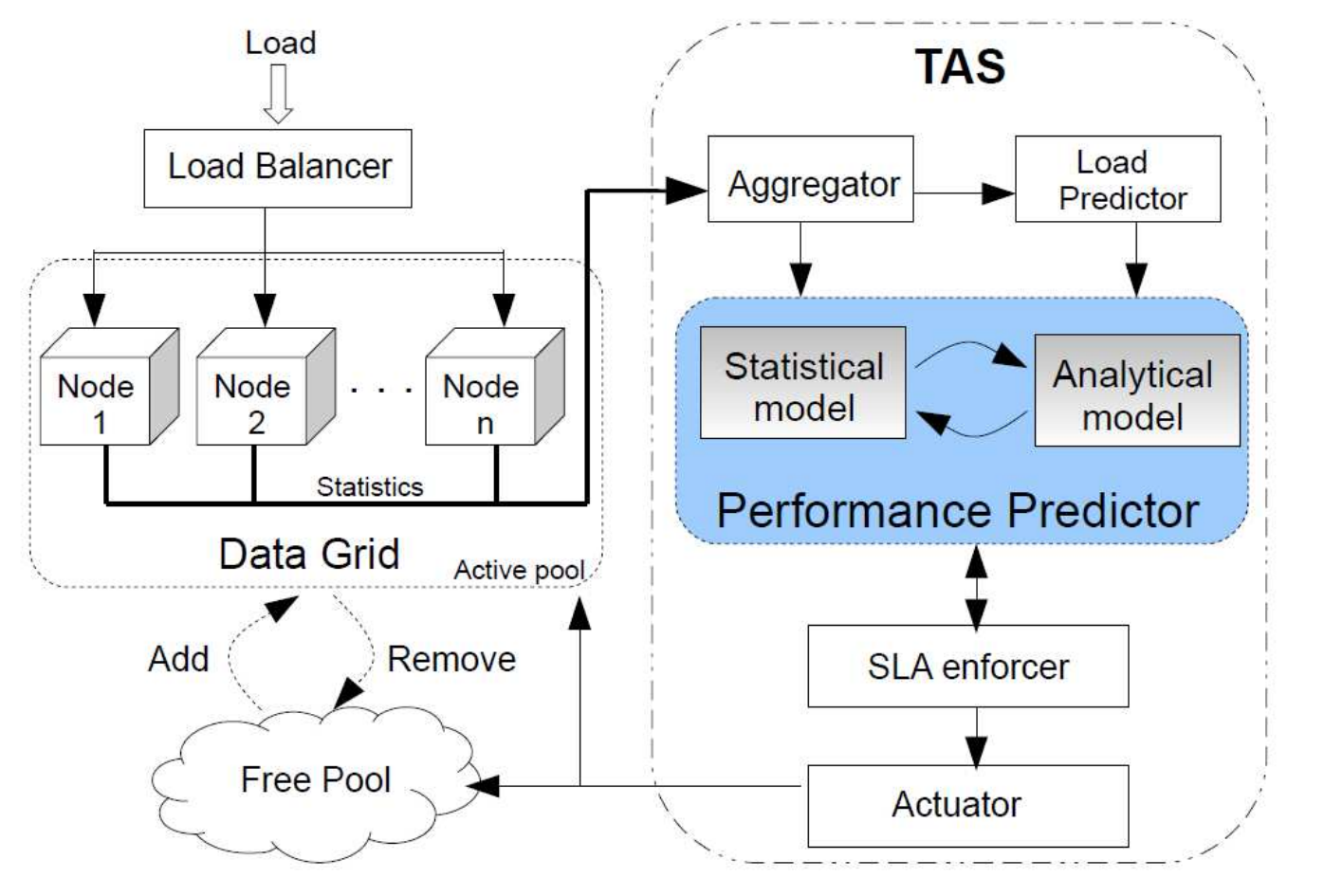}
\caption{TAS is a tool to help a tenant determine how many grid nodes
are needed to meet the demand \ElasticScaling.}
\label{fig:ElasticScaling}
\end{figure}

But first, why model a system?
To give an example, cloud providers offer their customers 
an attractive proposition that the resources provided can be elastically scaled
according to demand.
However, a tenant may need a tool to help determine how much 
resources (compute power, memory, etc.) to acquire,
but is hampered by having very little knowledge of the cloud architecture.
\ElasticScaling\
considers this problem in the context of a NoSQL data grid,
where the number of grid nodes can be increased or decreased
to match the workload
(see Figure~\ref{fig:ElasticScaling}). 
The authors present a tool (TAS) that uses equations to analytically model
how transaction throughput varies nonlinearly (and nonmonotonically,
sometimes) as the number of nodes increase.

\begin{figure}[t]
\centering
\includegraphics[height=5.8cm]{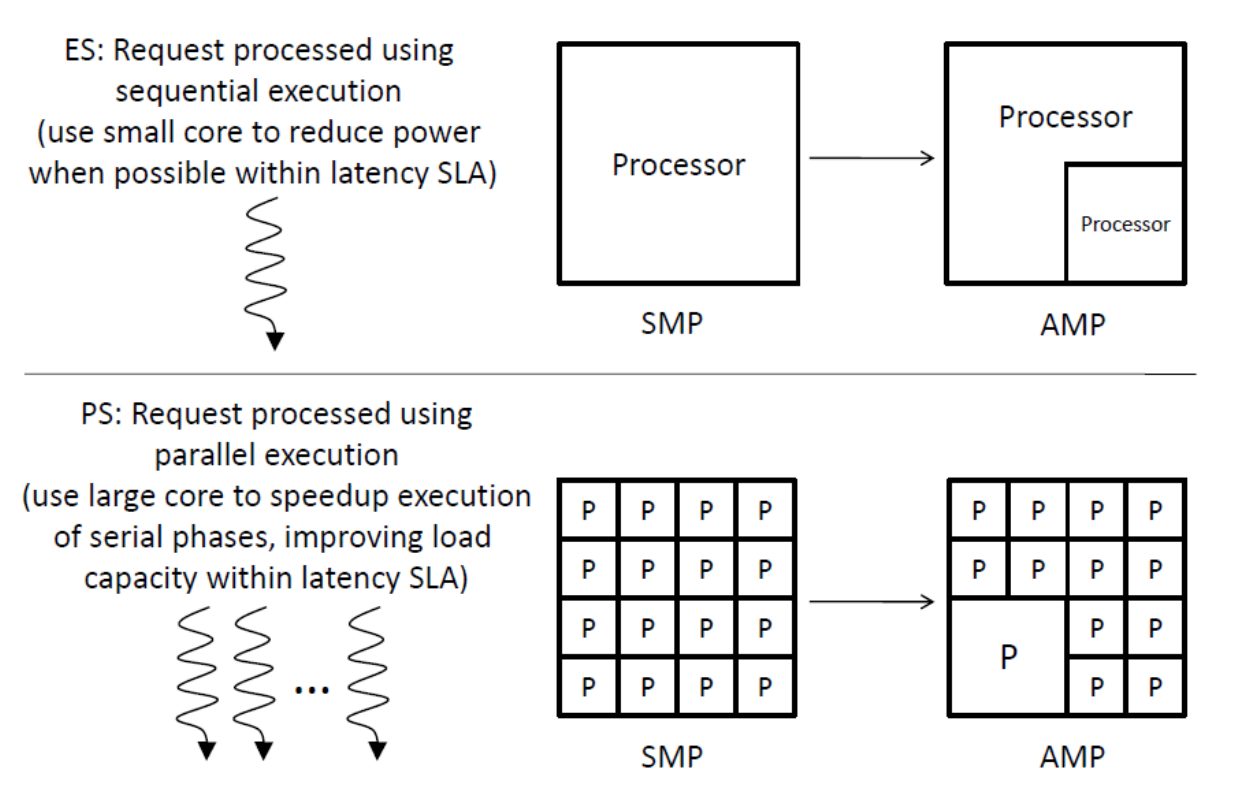}
\caption{Scaling a processor down or up 
(by reducing or increasing its number of cores)
to save energy or speed up computation \DatacenterAMP.}
\label{fig:DatacenterAMP}
\end{figure}

Energy and latency are major issues for data centers.
\DatacenterAMP\ uses a simple queueing model to study how asymmetric multicore
processors (AMP) can facilitate a tradeoff between energy use 
and latency bounds.
The idea is to dynamically marshal the resources of multiple cores
to give a larger processor that can speed up a sequential computation or,
when possible, scale it back down to reduce energy consumption
(see Figure~\ref{fig:DatacenterAMP}).

\GPU\ also uses a model for architectural exploration.
A GPU runs multiple threads simultaneously.
This speeds up the execution, but also causes delays from competition
for MSHRs (Miss Status Handling Registers) and the DRAM bus.
The paper uses a model to analyze how changing the number of MSHRs,
cache line size, miss latency, DRAM bandwidth, etc. affects the tradeoff.
There is no practical way of doing such an exploration with real hardware.
Part of the analysis is based on using the model to
generate the CPI (cycles per instruction)
stack that classifies the stall cycles into MSHR queueing delay, 
DRAM access latency, L1 cache hits, etc.

The model in \SoftErrors\
is similarly used to determine how various
microarchitectural structures (reorder buffer, issue queue, etc.)
contribute to the soft error rate that is induced by cosmic radiation.
Again, tweaking hardware structures is infeasible, 
and simulating soft errors is intractable,
so an analysis with a mathematical model is essential.

One equation from the \SoftErrors\ model explains why an intuition ---
that a workload with a low CPI should have a low vulnerability to soft errors
--- can be contradicted.
Similarly, the CPI stack breakdown by the \GPU\ model
explains how workload that spends many cycles queueing for DRAM bandwidth can, 
counter-intuitively, have negligible stall cycles from L2 cache misses.

To design a complicated system, an engineer needs help from intuition
that is distilled from experience.
However, experience with real systems is limited by availability and 
configuration history.
Although one can get around that via simulation,
the overwhelming size of the parameter space usually requires a limited
exploration; this exploration is, in turn, guided by intuition.
One way of breaking this circularity is to construct an analytical model
that abstracts away the technical details and zooms in on the factors
affecting the central issues.
We can then check our intuition with an analysis of the model.

Intuition is improved through contradictions: 
they point out limits on the old intuition, 
and new intuition is gained to replace the contradictions.
Again, such contradictions can be hard to find in a large simulation space,
but may be plain to see with the equations in a mathematical model.

The above examples illustrate the role that analytical models can play
in engineering complex computer systems.

\section{Assumptions}
\label{sec:Assumptions}

Analytical models are often dismissed 
because they are based on unrealistic assumptions.
For example, the latency model in \DatacenterAMP\ uses a simple $M/M/1$ queue
that assumes job inter-arrival and service times are exponentially distributed;
and the power model is a modification of Amdahl's Law \HM.
The exponential distribution has a strong ``memoryless'' property
and Amdahl's Law is an idealized program behavior.
Therefore, one expects the engineering complexity and overheads
of dynamically resizing a multicore processor will render any model
(simulators included) inaccurate.
However, the contribution of an analytical model often does not necessarily
lie in numerical accuracy, but possibly in providing insight 
into system behavior.
In the case of \DatacenterAMP,
its simple model suffices to reveal an instructive interplay among 
chip area, latency bound and power consumption.

Anyway, many models adopt equations from theory 
while breaking the assumptions that were used to derive those equations.
In \GPU, the model puts a bound on the delay computed with the 
Pollaczek-Khinchin formula for an $M/D/1$ queue.
One might consider this as a violation of the assumptions for the formula,
but another way to see it is that the bound merely modifies an approximation.
After all, the $M/D/1$ queue is itself an approximate model of DRAM
bandwidth contention delay.

A major result in queueing theory is Mean Value Analysis (MVA),
which derives an algorithm for calculating the performance of a 
{\bf separable} queueing network.
\DatabaseScalability\ shows how, by first profiling the performance of a 
standalone database, the MVA algorithm can be used to predict the performance
of a replicated database and, in addition, compare two alternative designs
(multi-master and single-master).
However, the probability of aborting a transaction increases with the
replication and the number of clients, thus changing the demand for resources
(at the CPU, disks, etc.) and thus breaking the MVA assumptions.
Yet, experiments show good agreement between model predictions
and simulation experiments.

\begin{figure}[t]
\centering
\includegraphics[height=3.5cm]{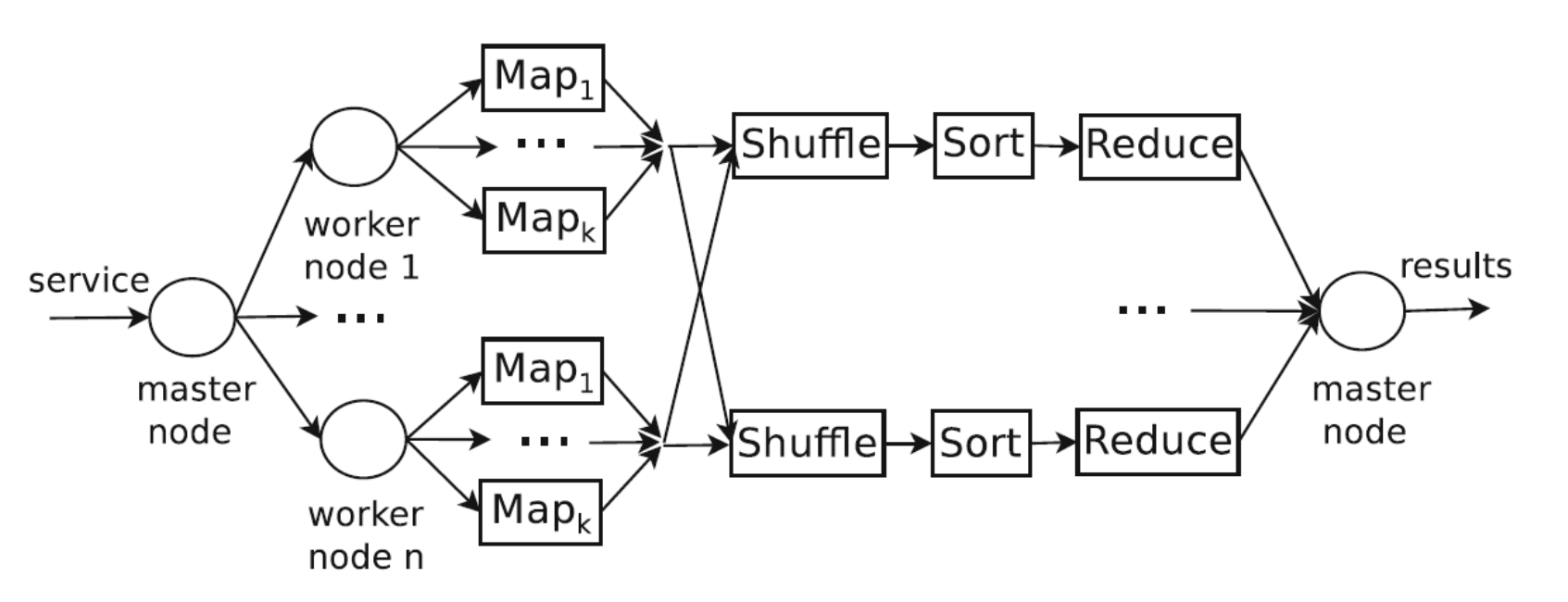}
\caption{
Precedence constraints among tasks in a MapReduce job adds to delays.
\MapReduce.}
\label{fig:MapReduce}
\end{figure}

\MapReduce\ is another example where the model uses the MVA algorithm
for a system that violates MVA assumptions.
Jobs in a system suffer delays for various reasons.
In this example, a job consists of {\it map}, {\it shuffle} and {\it reduce}
tasks that are delayed not only by queueing for resources,
but also by precedence constraints among the tasks 
(see Figure~\ref{fig:MapReduce}).
The key equation in the model is
\begin{equation}
\label{eq:MapReduce}
A_{ik}(\overrightarrow{N})=\sum_j f_{ij} Q_{jk}(\overrightarrow{N-1_i}),
\end{equation}
where $A_{ik}(\overrightarrow{N})$ and $Q_{jk}(\overrightarrow{N-1_i})$ 
are average queue lengths,
and $f_{ij}$ is a quantity determined by the precedence constraint;
this equation is then used in the MVA algorithm to iterate from job population
$\overrightarrow{N-1_i}$ to $\overrightarrow{N}$,
and thus solve the model.
Strictly speaking, however, precedence constraints violate MVA's separability
assumptions, and there is no $f_{ij}$ in Mean Value Analysis.

This practice of using equations even when their underlying assumptions
are violated recalls the proverbial reminder to keep the baby
(e.g. the Pollazcek-Khinchin formula, the MVA algorithm)
while throwing out the bath water (i.e. the assumptions).
This seems like a mathematical sin, but only if one confuses the sufficiency
and necessity of the equations' assumptions:
the equations may in fact be robust with respect to 
violations in their assumptions.

\section{Average Value Approximation (AVA)}
\label{sec:AVA}

The assumptions in an analytical model are often there to help justify
an equation.
\TCP\ is a well-cited mathematical analysis of the protocol.
The first step in its derivation for calculating expected throughput was
\[
E\left(\frac{Y}{A}\right) = \frac{EY}{EA},
\]
where $Y$ is the number of packets sent in a period $A$
between triple-duplicate packet loss indications.
This equation can be justified by assuming the TCP window size 
is a {\it Markov regenerative process},
but why should {\it this} assumption hold?

TCP has many states ({\it slow start}, {\it exponential backoff}, etc.)
and correlated variables (window size, timeout threshold, etc.).
These make it increasingly difficult in the analysis for one to 
even identify the assumptions needed to go from one equation to another.
Eventually, the authors simply adopt the approximation
\[
EQ(W)\approx Q(EW),
\]
without stating the assumptions.
Here, $Q(W)$ is the probability that 
a packet loss in a window of size $W$ is caused by a timeout,
$W$ is a random variable,
and the approximation estimates the average $Q(W)$ by $Q(EW)$,
i.e. replacing the variable $W$ with its average $EW$.
This technique, which I call {\bf Average Value Approximation (AVA)}~\Tay,
is widely used in performance modeling.

For example, \SoftErrors\ estimate the time an instruction stays in a 
reorder buffer as $\ell K$, where $\ell$ is the average latency per instruction,
and $K$ is the average critical path length.  
Strictly speaking, this time is $E(T_1+\cdots+T_n)$,
where $T_i$ is latency for instruction $i$,
$n$ is the critical path length,
$ET_i=\ell$ and $En=K$.
It is mathematically wrong to say $E(T_1+\cdots+T_n)=ET_1+\cdots+ET_n$,
since $n$ is a random variable.
If we use AVA and replace $n$ by $En$,
$T_{En}$ will not make sense if $E_n$ is not an integer;
if it is, then indeed $ET_1+\cdots+ET_{En} = \ell+\cdots+\ell=\ell K$.

A similar approximation appears in \fiveG,
which is a performance analysis of 5G cellular networks.
The system in such a network consists of overlapping wireless cells,
with bit rates that differ from cell to cell,
and from zone to zone within a cell.
How long would it take to download a file (e.g. movie)
as a user moves across the cells?
The final equation in the analysis estimates the average time as
\begin{equation}
\label{eq:5G}
\overline{T}=\overline{n_c}\overline{R}+\overline{n_h}t_h,
\end{equation}
where $\overline{n_c}$ is the average number of cells visited during the 
download, $\overline{R}$ is the average time the user spends in a cell,
$\overline{n_h}$ is the average number of handovers between cells,
and $t_h$ is the handover delay.
Notice the expression $\overline{n_c}\overline{R}$ is again an approximation
for $E(R_1+\cdots+R_{n_c})$,
where $R_i$ is the time spent in the i-th visited cell.

One could state some assumptions to rigorously justify
$E(T_1+\cdots+T_n)=\ell K$ and 
$E(R_1+\cdots+R_{n_c})=\overline{n_c}\overline{R}$,
but how does one justify those assumptions?
In analytical modeling of a complicated system,
one is focused on making progress with the derivation
(often liberally replacing random variables by their averages,
like $E\left(\frac{Y}{A}\right)=\frac{EY}{EA}$)
without worrying about the assumptions.
Whether the approximations have gone overboard will eventually be decided
by experimental validation.

\section{Bottleneck Analysis}
\label{sec:Bottleneck}

\begin{figure}[t]
\centering
\includegraphics[height=4.5cm]{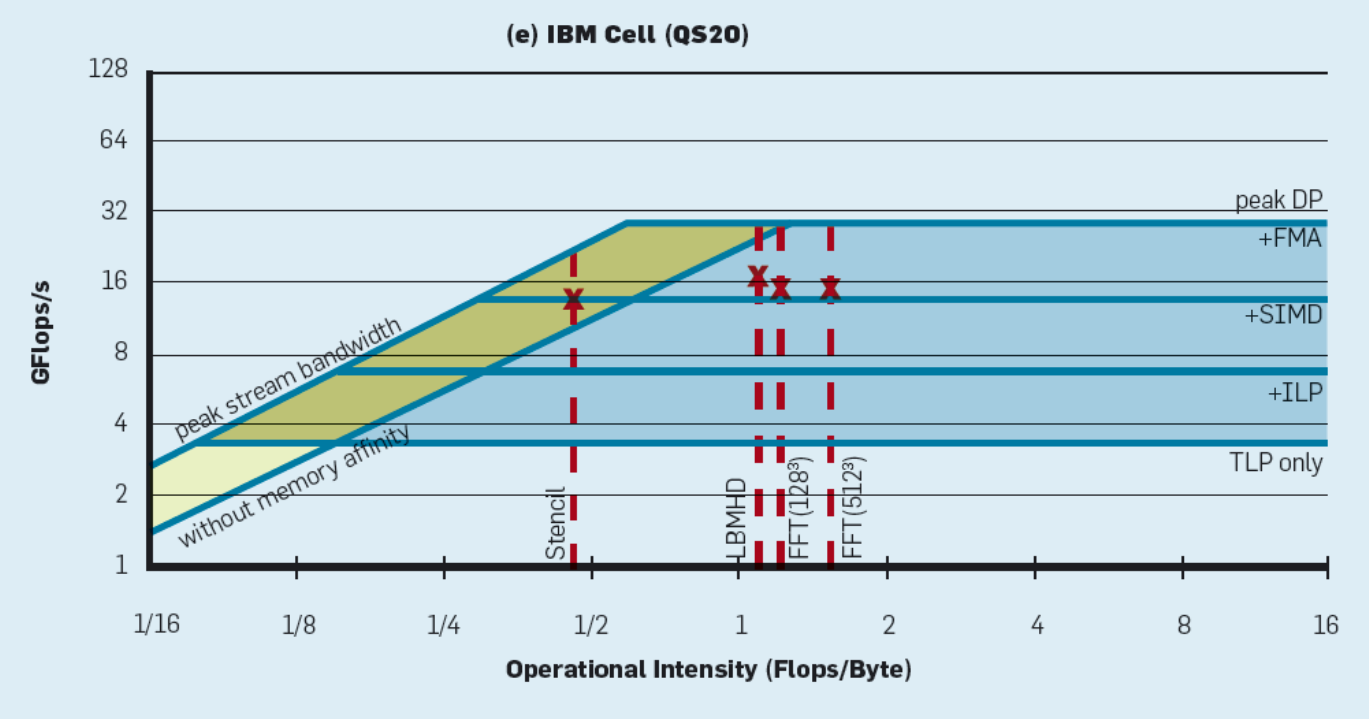}
\caption{
An application's rate of execution is bounded by memory bandwidth (sloping line)
and processor speed (horizontal line);
here, the application {\it stencil} is limited by memory bandwidth,
while {\it FFT} is limited by processor speed \Roofline.}
\label{fig:Roofline}
\end{figure}

There is a common misunderstanding that analytical modeling requires
queueing theory.
This is certainly untrue for the influential \Roofline\ model
for analyzing multicore architectures.
It is  a very simple model that has a {\bf roofline} 
with two straight segments, one representing the peak memory bandwidth
between processor caches and DRAM, and the other representing peak
floating-point processor performance (see Figure~\ref{fig:Roofline}).
No queueing theory is involved.

Despite its simplicity, the roofline model suffices for answering
some questions:
(i) Given an architecture and a floating-point kernel,
is performance constrained by processor speed or memory bandwidth?
The answer lies in comparing the kernel's {\it operational intensity}
(i.e. operations per byte of DRAM traffic)
to the two roofline segments.
For example, Figure~\ref{fig:Roofline} shows that the kernel {\it stencil} 
is limited by memory bandwidth, not processor speed.
(ii) Given a particular kernel, 
which architecture will provide the best performance?
One can answer this by comparing the rooflines of alternative architectures.
(iii) Given a particular architecture, how can the kernels be optimized 
to push the performance?
The optimization can be evaluated by examining how it shifts the roofline
and changes the operational intensity.

\Roofline\ makes very few assumptions and treats the processor as a blackbox.
Details like multithreading, how threads are distributed over the cores,
the cache hierarchy, etc. are not explicitly modeled.
Rather, it focuses on the possible bottlenecks
(memory bandwidth and processor speed) in the blackbox.

\begin{figure*}[t]
\centering
\includegraphics[height=5.8cm]{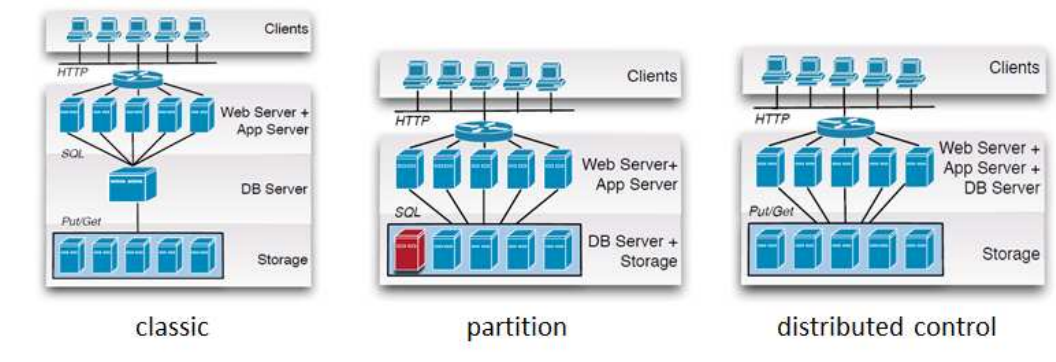}
\caption{
Three different cloud architectures for supporting transactions
\CloudTransactions.}
\label{fig:CloudTransactions}
\end{figure*}

Processor architecture aside, every blackbox has its bottlenecks.
This is true of datacenter-scale architectures as well.
For example, different cloud providers may have different architectures
for running transaction workloads.
\CloudTransactions\ consider three of these 
(see Figure~\ref{fig:CloudTransactions}).
What determines their scalability limits?

Once again, a cloud architecture is essentially a blackbox, 
with very little information 
on how the virtualized system is physically deployed.
Nonetheless, one can do a bottleneck analysis~\Tay\
to get the limits on transaction throughput:
\begin{eqnarray}
\lambda_{classic} &=& 
\min\left\{\frac{N_{WA}}{D_{WA}},\frac{c}{D_{db}},\frac{N_{st}}{D_{st}}\right\},
                                                                   \nonumber \\
\lambda_{partition} &=& 
\min\left\{\frac{N_{WA}}{D_{WA}},\frac{N^\prime_{st}}{D_{db}+D_{st}}\right\}, 
                                                                   \nonumber \\
\lambda_{dist.control} &=& 
\min\left\{\frac{N^\prime_{WA}}{D_{db}+D_{WA}},\frac{N_{st}}{D_{st}}\right\},
                                                                   \nonumber \\
\label{eq:CloudTransactions}
\end{eqnarray}
where {\it classic}, {\it partition} and {\it dist.control}
denote the three architectures in Figure~\ref{fig:CloudTransactions},
$N^\prime_{WA}$ and $N_{WA}$ are the number of web/application servers
with and without (respectively) co-located database servers;
$N^\prime_{st}$ and $N_{st}$ are the number of storage servers
with and without (respectively) co-located database servers;
$D_{WA}$, $D_{db}$ and $D_{st}$ are the service demands 
(seconds on a commodity machine ${\cal M}$) per transaction at a 
web/application, database and storage server (respectively), 
and $c$ is the ratio of database server speed in {\it classic} architecture 
to ${\cal M}$'s speed.

By scrutinizing the expressions in Equation~(\ref{eq:CloudTransactions}),
one can compare the scalability limits of two architectures,
and determine how one can push that limit for a particular architecture.
We thus see from Equation~(\ref{eq:CloudTransactions})
how an analytical model can bring clarity to the complex engineering choices
that support cloud computing.

\section{Parameter Space}
\label{sec:Parameters}

The experimental results in \CloudTransactions\ appear to show 
throughput increasing almost linearly for {\it dist.control}
as the number of emulated browsers $EB$ increases.
However, no system can scale its throughput linearly forever.
A closer examination of Equation~(\ref{eq:CloudTransactions})
shows that, for the parameter values in the experiments,
saturation should set in at around $EB=13500$,
but that is beyond the range of the experiments ($EB<9000$).
An analytical model can thus point out some important behavior that is beyond
the region of parameter space that is explored with experiments.

This advantage of an analytical model over experimental measurements 
is well-illustrated by \PPVoD,
which analyzes peer-to-peer (P2P) video-on-demand.
The viral success of P2P protocols inspired many ideas from academia 
for improving such systems. 
Each proposal can be viewed as a point in the design space,
whereas \PPVoD\ uses an analytical model to represent the entire space
spanned by throughput (number of bytes downloaded per second),
sequentiality (whether video chunks arrive in playback order)
and robustness (with respect to peer arrival/departure/failure,
bandwidth heterogeneity, etc.).
This global view of the parameter space leads to a counter-intuitive
observation, that a reduction in sequentiality can increase 
(sequential) throughput, thus demonstrating the point in Section~\ref{sec:Why}
that an analytical model can help improve intuition.

Furthermore, this model yields a {\it Tradeoff Theorem} that says 
a P2P video-on-demand system cannot simultaneously maximize throughput,
sequentiality and robustness.
This theorem is an example of how an analytical model can discover
the science that underlies an engineered system.

While an analytical model can give us a global view of the parameter space,
not all of this space is of practical interest.
Rather than delimit this space with some ``magic constant''
(e.g. $EB<9000$ in \CloudTransactions), \TransactionalMemory\ illustrates
how the restriction can be done through an aggregated parameter.

Single-processor techniques for coordinating access to shared memory
(e.g. semaphores) do not scale well as the number of cores increases,
and \TransactionalMemory\ examines an alternative.
For the software transactional memory in this study,
locks are used to control access to shared objects,
and locks in a piece of code are grouped into transactions;
each transaction is, in effect, executed atomically.
The parameter space in the experiments is restricted to $\frac{kN}{L}<1$,
where $k$ is the number of locks per transaction,
$N$ is the number of concurrent transactions,
and $L$ is the number of objects that can be locked
(so, intuitively, the system is overloaded if $kN>L$).

In fact, one can show that the performance (throughput, abort rate, etc.)
is determined by $k$ and $\Lambda$,
where $\Lambda=\frac{N}{L}$ \TGS,
thus reducing the parameter space from 3-dimensional ($k, N, L$) 
to 2-dimensional ($k, \Lambda$).
The use of such aggregated parameters ($\frac{kN}{L}$ and $\Lambda$)
significantly reduces the space that the experiments must cover.

\begin{figure}[t]
\centering
\includegraphics[height=6.1cm]{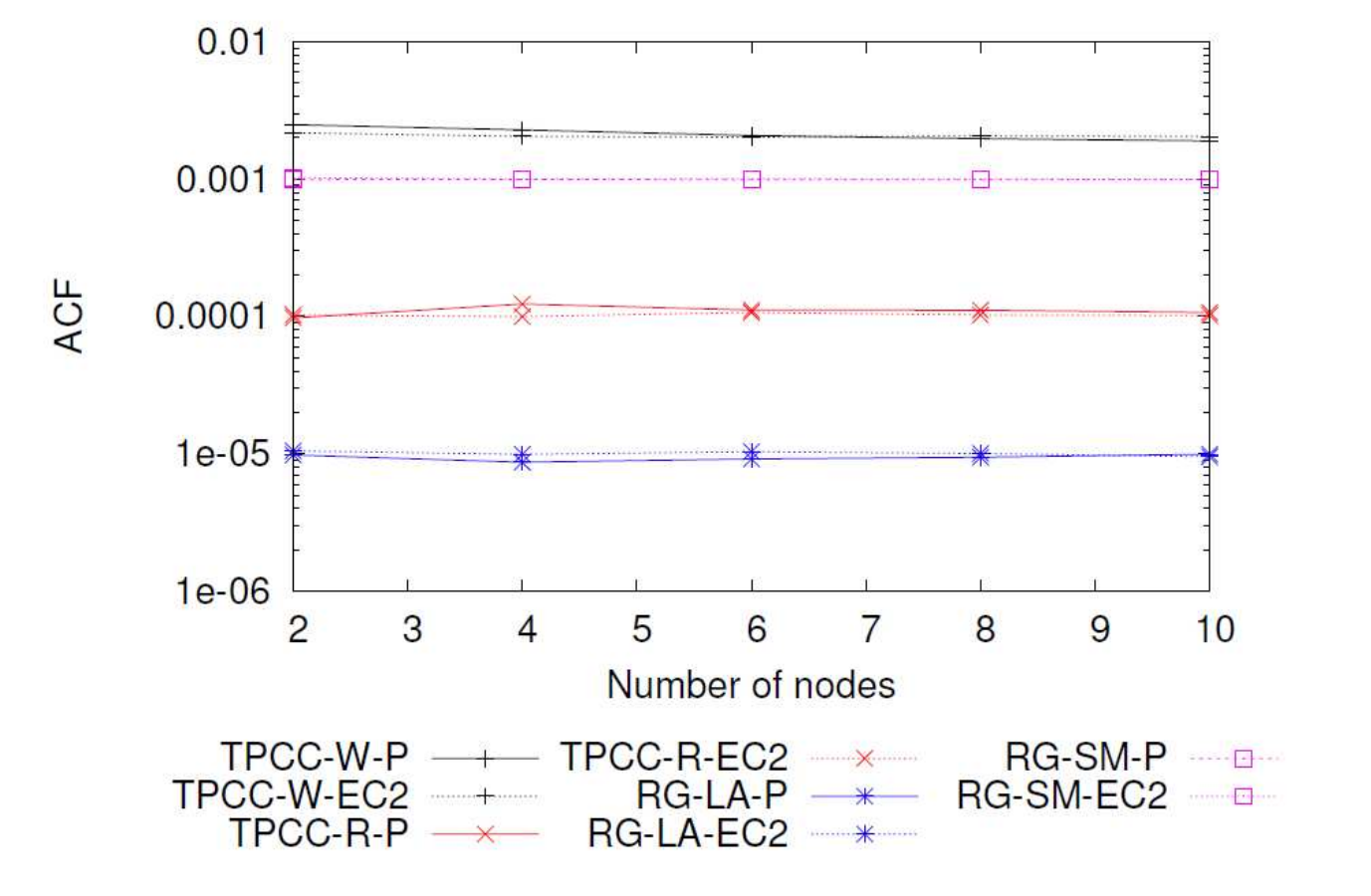}
\caption{
The Application Contention Factor $ACF$ 
is a property of the transactions;
it does not vary with the number of grid nodes,
and is the same for private and public clouds \ElasticScaling.
}
\label{fig:ACF}
\end{figure}

Parameter aggregation can go much further.
One key issue in analyzing transaction performance 
is modeling the access pattern.
Most studies assume access is {\it uniform},
i.e. the probability of requesting any one of $L$ objects is $\frac{1}{L}$.
In reality, some objects are more frequently requested than others,
but how many parameters would it take to model a realistic access distribution?
One solution to this difficulty was discovered by \ElasticScaling.
This paper defines an {\it Application Contention Factor} 
$ACF=\frac{P_{\rm lock}}{{\lambda_{\rm lock}}{T_{\rm hold}}}$,
where $P_{\rm lock}$ is the probability of a lock conflict,
$\lambda_{\rm lock}$ is the rate of lock requests and
$T_{\rm hold}$ is the average lock holding time.
Experiments show that the $ACF$ of a given workload is constant
with respect to the number of nodes in the data grid,
and the same whether it is run in a private or public cloud
(see Figure~\ref{fig:ACF}).

Now, let $D=\frac{1}{ACF}$, so 
\begin{equation}
P_{\rm lock}=\frac{\lambda_{\rm lock}T_{\rm hold}}{D}.
\label{eq:ACF}
\end{equation}
By Little's Law, $\lambda_{\rm lock}T_{\rm hold}$ 
is the expected number of objects that are locked,
so Equation~(\ref{eq:ACF}) says the probability of conflict is as if 
access is {\it uniformly} distributed over $D$ objects.
This is because the access pattern, being a property of the transactions,
should be constant with respect to the number of grid nodes,
and the same whether the cloud is private or public.

This observation simplifies tremendously the analysis of transaction behavior.
It says the large literature that rests on the assumption of uniform access 
remains valid once the number of objects is reinterpreted via $ACF$;
going forward, we can continue to adopt uniform access in
modeling transaction performance.
This is another example where analytical modeling is helping to discover
the science in engineering computer systems.

\section{Decomposition and Decoupling}
\label{sec:Decomposition}

Analyzing a computer system is not any easier if it is as small as a chip,
as is the case with \SoftErrors\ (see Section~\ref{sec:Why}).
The probability that a radiation-induced fault in a microarchitecture
structure (e.g. reorder buffer or issue queue)
causes a program output error depends on both the hardware (microarchitecture)
and the software (program).
Recall from Section~\ref{sec:AVA} that the model estimates the time
an instruction stays in a reorder buffer as $\ell K$,
where $\ell$ is the average instruction latency,
and $K$ is the average critical path length.
It thus decouples the hardware and software:
one can analyze separately the impact of changing the hardware 
(which determines $\ell$) and changing the workload (which determines $K$).
For example, we can evaluate how changing issue width affects performance,
without having to re-profile the workload.
Such decoupling is a powerful technique in the scientific analysis of
a complicated system.

One can view such a decoupling as decomposing into two submodels:
one for hardware, the other for software.
\MapReduce\ has a similar decomposition.
In Equation~(\ref{eq:MapReduce}),
$f_{ij}$ is a factor determined by the precedence constraint,
while $A_{ik}$ and $Q_{jk}$ are queue lengths in the queueing network.
The precedence constraint is a model for the job execution,
while the queueing network is a model for the resource contention.
These two submodels are not completely decoupled,
as evaluating $f_{ij}$, $A_{ik}$ and $Q_{jk}$ requires an iteration
between the graph and queueing models.

There is also model decomposition in \ElasticScaling.
Figure~\ref{fig:ElasticScaling} shows the performance predictor consists of
two submodels:
(i) A statistical model for the resource contention;
since the resources are virtualized and the hardware configuration is unknown,
this model uses machine learning to model how the cloud performance responds
to changes in the workload.
(ii) An analytical model for the data contention, 
i.e. how the number of locks, transactions and objects affect
the probability that a transaction encounters a lock conflict,
or has to abort.
We see here that the blackbox/whitebox difference mentioned in 
Section~\ref{sec:Intro} does not mean we must choose just one of them.
The \ElasticScaling\ has both:
the machine learning model is a blackbox,
while the analytical model is a whitebox;
the two are developed independently but, again,
the performance prediction requires an iteration between the two.

\section{Analytic Validation}
\label{sec:AnalyticValidation}

\begin{figure}[t]
\centering
\includegraphics[height=6.1cm]{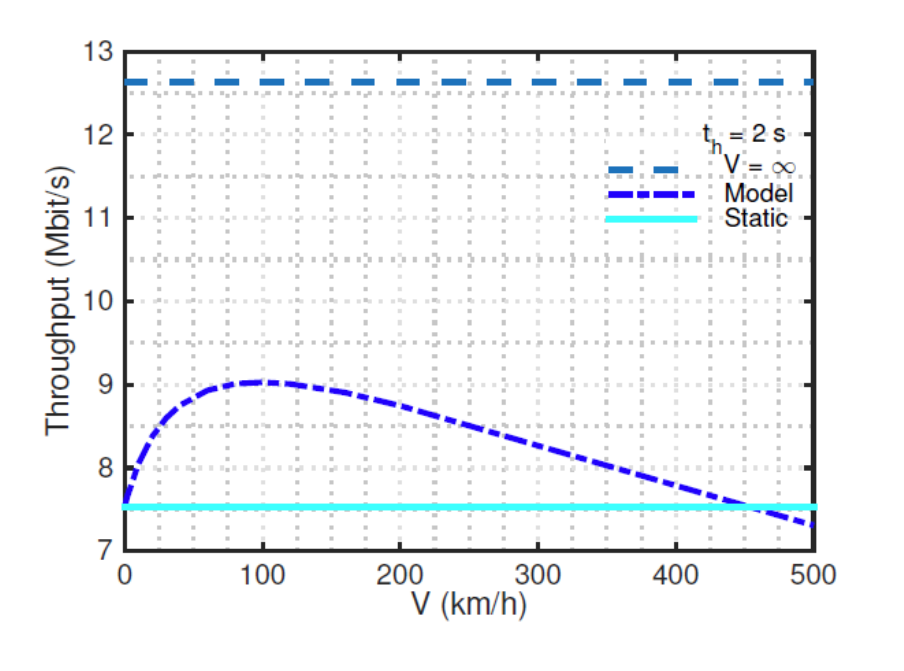}
\caption{
As user speed increases while traversing 5G cells,
the throughput for a file download at first increases, then decreases
\fiveG.
}
\label{fig:5G}
\end{figure}

Although analytical models are often touted as an alternative to simulators,
many of them are actually used as simulators.
To give an example, \fiveG\ presents an interesting plot showing that,
as user speed increases, throughput at first increases, then decreases
(see Figure~\ref{fig:5G}).
This plot was not generated by a simulator,
but by numerical solution of the \fiveG\ analytical model;
and the nonmonotonic throughput behavior is observed from the plot,
not proved with the model.
In this sense, the model is used as a simulator;
I call this {\bf analytic simulation} \Tay.

If we draw conclusions from plotting numerical solutions of an analytical model,
how do we know if these conclusions are about the model,
rather than the system?
This is a possibility because the model is only an approximation of the system
(this is an issue for simulators as well).

To give an example, classical MVA solutions of a queueing network model
has throughput that increases monotonically as the number of jobs increases.
If such a model were used for \ElasticScaling, 
it will fail to capture a critical behavior (throughput nonmonotonicity)
in the system (see Section~\ref{sec:Why}).
Indeed, one can find examples in the literature of such a qualitative 
divergence in properties between model and system.

We should therefore verify that an interesting property observed in a model is,
in fact, a property of the system.
I call this {\bf analytic validation} \Tay.

For example, the \ElasticScaling\ model claims that Application Contention
Factor $ACF$ is a constant that is a property of the transactions
(independent of the number of grid nodes and whether the transactions
run in a private or public cloud),
and Figure~\ref{fig:ACF} presents an experiment to validate this claim.
Note that nothing in this plot is generated by the model ---
the properties we observe in the data are properties of the system itself.
In contrast, in a {\it numeric} validation of an analytical model,
there is always a comparison of experimental measurements 
to numerical solutions from the model.

We can see another example of analytic validation in \RouterBuffer.
Internet routers drop packets when their buffers are full.
To avoid this, and to accommodate the large number of flows,
router buffers have become very large.
\RouterBuffer\ uses an analytical model to study whether
flow multiplexing can make this buffer bloat unnecessary.
Among the results is an expression that says buffer occupancy $EQ$
depends on the link rate only through utilization $\rho$.
This is a strong claim from the model, 
so it requires experimental verification.

\begin{figure}[t]
\centering
\includegraphics[height=5.8cm]{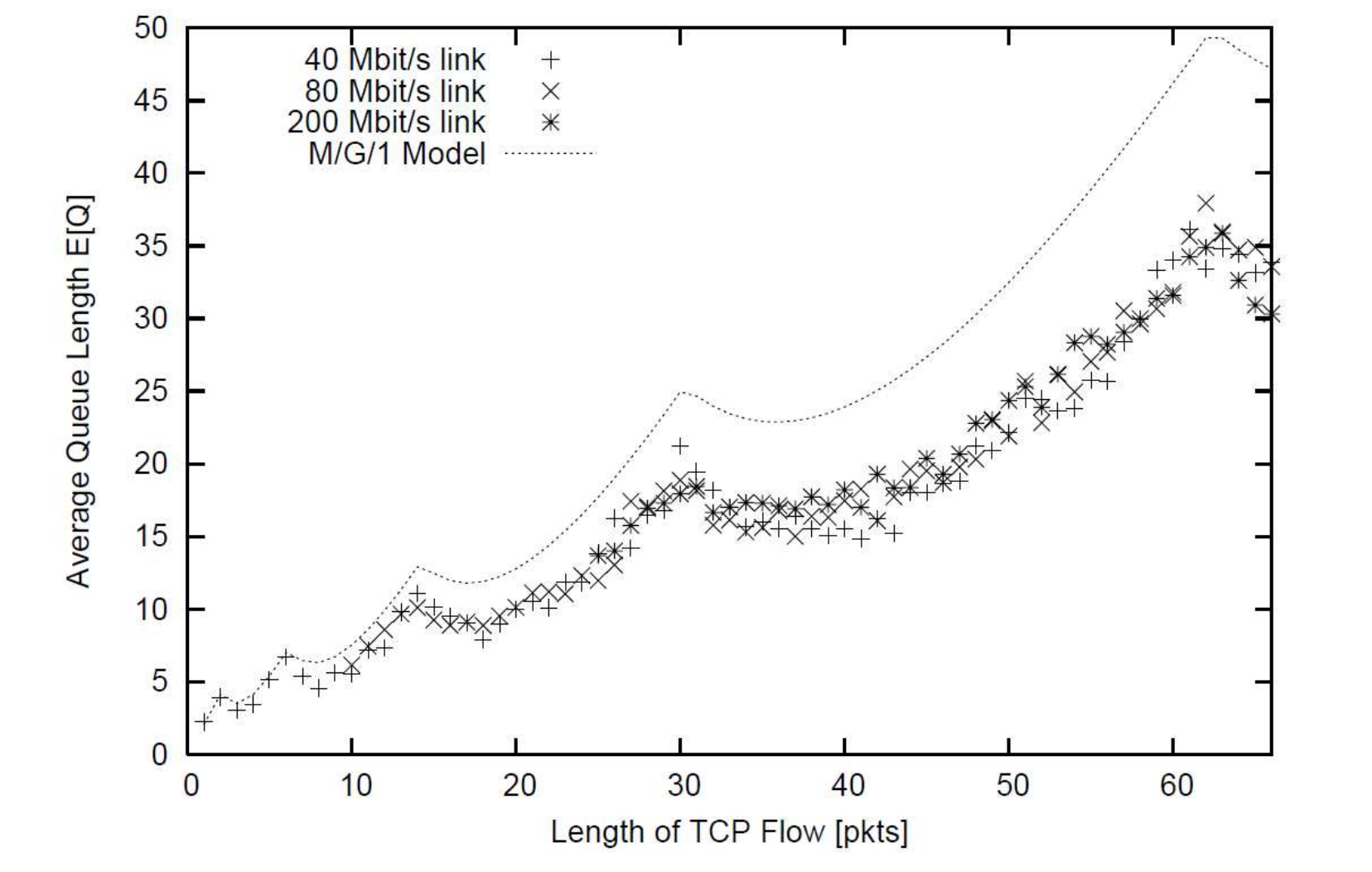}
\caption{
For a fixed utilization $\rho=0.8$,
simulated buffer occupancy $EQ$ for different link rates are similar.
Moreover, they have a saw-tooth behavior,
as predicted by the model \RouterBuffer.
}
\label{fig:RouterBuffer}
\end{figure}

Figure~\ref{fig:RouterBuffer} shows that, for fixed $\rho=0.8$,
simulation measurements of $EQ$ are indeed similar for different link rates.
Furthermore, a numerical solution of the model shows that $EQ$ 
has a saw-tooth behavior as the TCP flow length increases;
Figure~\ref{fig:RouterBuffer} shows that the simulated $EQ$
has this behavior too.
The numerical difference between model and simulation 
in the plot is beside the point;
rather, the experiment shows that an unusual behavior predicted by the model
is not an artifact of the assumptions and approximations,
but is in fact a property of the system itself.

For another illustration, consider the IEEE 802.11 protocol for WiFi:
it specifies what a base station and the mobile devices in its wireless cell
should do in sending and receiving packets.
Simultaneous transmissions from different mobile devices can cause
packet collisions and induce retransmission and backoff,
so maximum throughput in the cell can be lower than channel bandwidth.
The model in \WiFi\ examines how this saturation throughput depends on the
protocol parameters, and analyzes the tradeoff between collision and backoff.

\begin{figure*}[t]
\centering
\includegraphics[height=6.5cm]{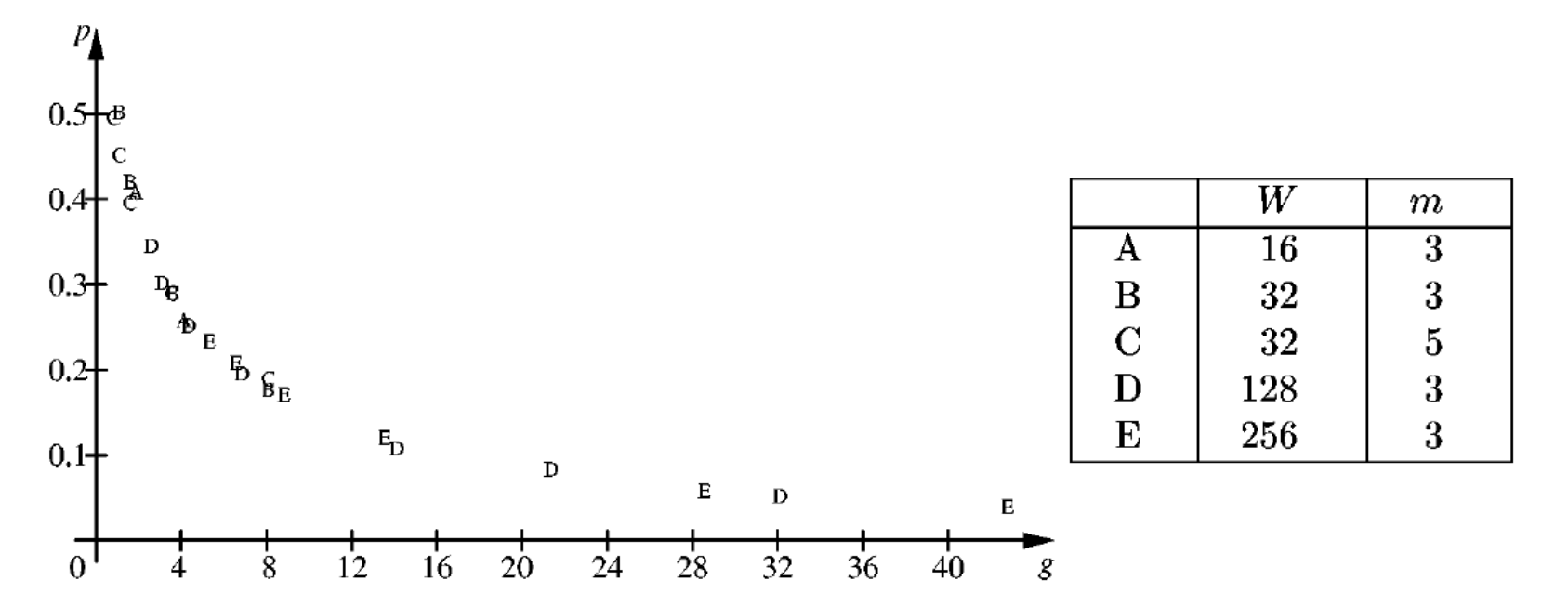}
\caption{
Probability of packet collision $p$ depends on minimum window size $W$
and number of mobile devices $n$ only through the gap $g=\frac{W}{n-1}$.
The maximum window size is $2^m W$. \WiFi
}
\label{fig:WiFi}
\end{figure*}

One of the claims from the \WiFi\ model says that the probability $p$
of a collision depends on the protocol's minimum window size $W$ 
and the number of mobile devices $n$ only through the 
{\it gap} $g=\frac{W}{n-1}$.
Figure~\ref{fig:WiFi} shows that $\langle g,p\rangle$ from simulations
with different configurations do in fact lie on the same curve.
As a corollary, this reduces the 3-parameter space $\langle n, W, m\rangle$
--- where maximum window size is $2^m W$ ---
to just a single parameter $g$ (recall Section~\ref{sec:Parameters}).

The model also claims that bandwidth wasted by packet collisions
exceeds idle bandwidth caused by backoffs if and only if $r>\frac{1}{T}$, 
where $r$ is the transmission rate and 
$T$ is the transmission time (including packet headers);
this was also analytically validated by the experiments.
Here, we see how an analytical model can discover the science
that governs wireless packet transmissions over a shared channel.

\section{Analysis with an Analytical Model}
\label{sec:Analysis}

In answering the question ``Why an Analytical Model?'' (Section~\ref{sec:Why}),
we give examples where the model is used for 
architectural exploration (\GPU\ and \SoftErrors) and
resource provisioning (\ElasticScaling); 
other examples include \DatacenterAMP, \CloudTransactions\ and \Roofline\
(design exploration) and \MapReduce\ (capacity planning).
For these examples, the analytical models were used to generate 
numerical predictions.

This analytic simulation can sometimes reveal interesting behavior
in a system (e.g. \fiveG\ in Section~\ref{sec:AnalyticValidation})
but such revelations could arguably be obtained by a more realistic,
albeit slower, simulation model.
The power in an analytical model lies not in its role as a fast substitute
for a simulator, but in the analysis that one can bring to bear 
on its equations.
Such an analysis can yield insights that cannot be obtained by eyeballing
any number of plots --- without knowing what you are looking for ---
and provide conclusions that no simulator can offer.

We see illustrations of this power in \ElasticScaling,
which discovers that nonuniform access is equivalent to uniform access
via the Application Contention Factor;
in \TransactionalMemory, which shows two dimensions of the parameter space
($N$ and $L$) can be reduced to one ($\frac{N}{L}$);
in \PPVoD, where the Tradeoff Theorem says throughput, sequentiality
and robustness cannot be simultaneously maximized for P2P video-on-demand;
in \WiFi, which characterizes the optimal balance between bandwidth loss
through packet collisions and time loss to transmission backoff.
There are two other examples in \RouterBuffer\  and \TCP.

\RouterBuffer\ shows that (i) for $n$ long TCP flows,
buffer size should be inversely proportional to $\sqrt{n}$ and
(ii) for short flows, the probability of a buffer overflow does not depend
on $n$, round-trip time, nor link capacity.
These are insights obtained from the analytical model
--- one cannot get them from any finite number of simulation experiments.

The key equation in \TCP\ expresses throughput in terms of packet loss 
probability $p$ and round-trip time $RTT$.
Clearly, for any nontrivial Internet path,
$p$ and $RTT$ cn only be measured, not predicted,
so what is the point of having that equation?
Its significance lies not in predicting throughput,
but in characterizing its relationship with $p$ and $RTT$.
Such a characterization led to the concept of TCP-friendliness and
the design of equation-based protocols.

\RouterBuffer\ and \TCP\ thus advance the science of network communication.

For a final, topical example, consider the spread of fake news and memes, etc.
over the Web, driven by user interest,
modulated by daily and weekly cycles, and dampened over time.
To adequately capture this behavior,
\InformationDiffusion\ modifies the classical epidemic model.
However, there is no way of integrating the resulting differential equation,
so it is solved numerically.
The parameters of the model can then be calibrated by 
fitting measured data points.
Since measurements are needed for calibration,
the model has limited predictive power.
Nonetheless, the parametric values serve to succinctly characterize
the diffusion, and provide some insight into its spread.
This example illustrates the point that,
although the pieces of a system are designed, engineered and artificial,
they can exhibit a hard-to-understand, organic behavior when put together.
An analytical model is thus a tool for developing the science of such
organisms.

(Incidentally, given the recent, tremendous success of 
artificial neural networks,
it is fashionable now to speculate that artificial intelligence
may one day become so smart that some AI system will control humans.
I believe that, if that day comes, the system will behave biologically,
with biological vulnerabilities that are open to attack.)

\section{Conclusion}
\label{sec:Conclusion}

This review has discussed some issues 
(Sec.~\ref{sec:Assumptions} and {Sec.~\ref{sec:AnalyticValidation})
and introduces some key ideas and techniques 
(Sec.~\ref{sec:AVA}, Sec.~\ref{sec:Bottleneck}, 
Sec.~\ref{sec:Parameters} and {Sec.~\ref{sec:Decomposition})
in analytical performance modeling.
Although the models are often conceived as an engineering tool to generate
numerical predictions for the design and control of computer systems
(Sec.~\ref{sec:Why}),
Sec.~\ref{sec:Analysis} points to a less obvious role
in discovering the science that underlies these systems, 
much like the role that mathematical models play
in discovering physics in nature.
This is a role that spans all computer systems 
(processor, memory, bandwidth, database and multimedia systems, etc.),
that helps develop a {\it Computer Science} that withstands 
changes in technology.

\section{References}

\begin{description}

\item[[5G]]
B. Baynat and N. Narcisse,
Performance model for 4G/5G networks taking into account 
intra- and inter-cell mobility of users.
{\it Proc. LCN 2016}, 212--215.

\item[[802.11]]
Y. C. Tay and K. C. Chua. 
A capacity analysis for the IEEE 802.11 MAC protocol.
{\it Wireless Networks}, 7(2):159--171 (2001).

\item[[CloudTransactions]]
D. Kossman, T. Kraska and S. Loesing.
An evaluation of alternative architectures for transaction processing 
in the cloud.
{\it Proc. SIGMOD 2010}, 579--590.

\item[[DatacenterAMP]]
V. Gupta and R. Nathuji.
Analyzing performance asymmetric multicore processors 
for latency sensitive datacenter applications.
{\it Proc. HotPower 2010}, 1--8.

\item[[DatabaseScalability]]
S. Elnikety, S. Dropsho, E. Cecchet and W. Zwaenepoel.
Predicting replicated database scalability from standalone database profiling.
{\it Proc. EuroSys 2009}, 303--316.

\item[[ElasticScaling]]
D. Didona, P. Romano, S. Peluso and F. Quaglia.
Transactional Auto Scaler:
Elastic scaling of in-memory transactional data grids.
{\it Proc. ICAC 2012}, 125--134.

\item[[GPU]]
J.-C. Huang, J.H. Lee, H. Kim and H.-H. S. Lee.
GPUMech: GPU performance modeling technique based on interval analysis.
{\it Proc. MICRO 2014}, 268--279.

\item[[HM]]
M.D. Hill and M.R. Marty.
Amdahl's Law in the Multicore Era.
{\it {IEEE} Computer},
41(7):33--38 (2008).

\item[[InformationDiffusion]]
Y. Matsubara, Y. Sakurai, B.A. Prakash, L. Li and C. Faloutsos.
Rise and fall patterns of information diffusion: model and implications.
{\it Proc. KDD 2012}, 6--14.

\item[[MapReduce]]
E. Vianna, G. Comarela, T. Pontes, J. Almeida, V. Almeida, K. Wilkinson, H. Kuno
 and U. Dayal.
Analytical performance models for {MapReduce} workloads.
{\it Int. J. Parallel Programming} 41(4):495--525, August 2013.

\item[[P2PVoD]]
B. Ran, D.~G. Andersen, M. Kaminsky and K. Papagiannaki.
Balancing throughput, robustness, and in-order delivery in P2P VoD.
{\it Proc. CoNEXT 2010}, 10:1--10:12.

\item[[PipelineParallelism]]
A. Navarro, R. Asenjo, Si. Tabik and C. Cascaval.
Analytical modeling of pipeline parallelism.
{\it Proc. PACT 2009}, 281--290.

\item[[Roofline]]
S. Williams, A. Waterman and D. Patterson.
Roofline: an insightful visual performance model for multicore architectures.
{\it CACM}, 65--76, April 2009.

\item[[RouterBuffer]]
G. Appenzeller, I. Keslassy and N. McKeown.
Sizing router buffers.
{\it Proc. SIGCOMM 2004}, 281--292.

\item[[SoftErrors]]
A.A. Nair, S. Eyerman, L. Eeckhout and L.K. John.
A first-order mechanistic model for architectural vulnerability factor.
{\it Proc. ISCA 2012}, 273--284.

\item[[T]]
Y.C. Tay, {\sl Analytical Performance Modeling for Computer Systems}
(3rd Edition), Morgan \& Claypool Publishers, 2018.

\item[[TCP]]
J. Padhye, V. Firoiu, D. Towsley, and J. Kurose. 
Modeling TCP throughput: a simple model and its empirical validation. 
{\it Proc. SIGCOMM 1998}, 303--314.

\item[[TGS]]
Y.C. Tay, N. Goodman and R. Suri.  
Locking performance in centralized databases.  
{\it ACM Transactions on Database Systems 10}, 4(Dec. 1985), 415--462.  

\item[[TransactionalMemory]]
A. Heindl, G. Pokam, and A.-R. Adl-Tabatabai. 
An analytic model of optimistic software transactional memory. 
{\it Proc. ISPASS 2009}, 153--162.

\end{description}
\end{document}